\documentclass{article}

\usepackage{arxiv}

\usepackage[utf8]{inputenc} % allow utf-8 input
\usepackage[T1]{fontenc}    % use 8-bit T1 fonts
\usepackage{hyperref}       % hyperlinks
\usepackage{url}            % simple URL typesetting
\usepackage{booktabs}       % professional-quality tables
\usepackage{amsfonts}       % blackboard math symbols
\usepackage{nicefrac}       % compact symbols for 1/2, etc.
\usepackage{microtype}      % microtypography
\usepackage{xcolor}
\usepackage[linesnumbered,ruled,vlined]{algorithm2e}
\usepackage[noend]{algpseudocode}
\usepackage{graphicx}
\usepackage{ulem}
\usepackage{subfigure}

\title{Deep learning for peptide identification from metaproteomics datasets}

\author{
Xuan Guo \\
Department of Computer Science and Engineering\\
University of North Texas\\
Denton, TX\\
\texttt{Xuan.Guo@unt.edu}\\
\And
Shichao Feng\\
Department of Computer Science and Engineering\\
University of North Texas\\
Denton, TX\\
\texttt{FengFeng@my.unt.edu}\\
}

\SetCommentSty{mycommfont}

\SetKwInput{KwInput}{Input}                % Set the Input
\SetKwInput{KwOutput}{Output}              % set the Output

\begin{document}
\maketitle
\begin{abstract}
Metaproteomics are becoming widely used in microbiome research for gaining insights into the functional state of the microbial community. Current metaproteomics studies are generally based on high-throughput tandem mass spectrometry (MS/MS) coupled with liquid chromatography. The identification of peptides and proteins from MS data involves the computational procedure of searching MS/MS spectra against a predefined protein sequence database and assigning top-scored peptides to spectra. Existing computational tools are still far from being able to extract all the information out of large MS/MS datasets acquired from metaproteome samples. In this paper, we proposed a deep-learning-based algorithm, called DeepFilter, for improving the rate of confident peptide identifications from a collection of tandem mass spectra. Compared with other post-processing tools, including Percolator, Q-ranker, PeptideProphet, and Iprophet, DeepFilter identified 20\% and 10\% more peptide-spectrum-matches and proteins, respectively, on marine microbial and soil microbial metaproteome samples with false discovery rate at 1\%. 

%Characterized from the mixture of all the proteins expressed by a complex microbial community, metaproteomics analysis has already regarded as a powerful method to investigate protein identification. However, due to the complexity of species, incomplete of database recording, and redundant peptide identification, intentional protein identification database searching algorithm has limitation to detect target peptides in metaproteomics. In this way, it requires a more efficient and effective post-processing tool to increase the peptide identification quality. To achieve this goal, we selected samples with high-quality from partial marine microbial communities and trained a post-processor DeepFilter based on convolution neural network (CNN) after Comet database searching engine. 
\end{abstract}

\section{Introduction}

Metaproteomics is the analysis of the protein samples from multi-organisms in a specific environment. Because of the significant role they play in nutrient cycling and the immune system, complex microbial communities studies have gained increasing attention in recent years. Metaproteomics analysis is mostly based on the shotgun proteomics technique which uses high-pressure liquid chromatography-tandem mass spectrometry (LC-MS/MS). 

The complex microbial community data sets are worth analyzing, but some features of them impede the MS-based metaproteomics. Directly derived from the environment, the large amount \cite{schluter2008metagenome} of microbial species causes complexity and heterogeneity for protein database-searching. As \cite{isaac2019metaproteomics} introduces, in a typical metaproteomics environment, there are more than 1000 unique species, and each one includes several hundred proteins, which causes a very massive amount of peptide sequences after digestion, so the requirement of computational effort and memory to store the digested peptides have to be increased. Besides, since the number of microbial species is far greater than the number of species recorded in the published protein databases \cite{locey2016scaling}, the task of protein identification is difficult when the taxonomic composition of protein expression is not recorded in the existing database \cite{heyer2017challenges}. Also, the homologous protein sequences digest into common peptide candidates, which cause redundant peptide identification. That is, the common peptide candidates will be scored with different mass spectra randomly; as the size of the protein database and the number of spectra increase, the probability of determining the selected false-positive samples with high-scoring PSM increases. In this way, the progress of shotgun proteomics enhances the resolution of tandem mass spectrometry, the challenge of metaproteomics itself, and the advance of technique make it thirsty to improve protein efficiency identification within MS-based data.

We propose a deep learning method to post-process the PSM candidates after the database-searching engine improves the quality and efficiency of protein identification for metaproteomics. We hope this method can help recognize the potential pattern of the mass spectrum to reduce false positive sample identification and increase protein identification. In section 2, we introduce the popular database-searching engines and post-processors. In this section, we also discuss some researchers' work showing the potential of using deep learning to improve protein identification performance. In the method section, we detail the whole workflow and architecture of our DeepFilter post-processing system, and state the baseline use. In the result section, we present the performance within complex microbial communities as well as single organisms, and compare our model with other baselines. Finally, we use  class activation mappings to visualize the features learned by the deep learning model.

\section{Related work}
Mass-spectrometry-based metaproteomics has become a typical and effective technique for protein identification. Following the basic principles, protein is digested, and the sequences from the database for specific microbial communities are transferred into predicted peptides. The process we use to compare and find the matching tandem mass spectrometry data and predict peptides is database searching. Database searching algorithms score these peptide-spectrum matches (PSMs) by comparing experimental mass spectrum data and peptides from the database using mathematical or statistic methods. Comet \cite{eng2013comet} uses cross-correlation to calculate the score between experimental peptide spectrum and theoretical peptide. Some other algorithms mostly regard probability-based scoring functions to identify the quality of PSMs: Myrimatch \cite{tabb2007myrimatch} uses a Multivariate hypergeometric distribution to get random matches and stratify the peak intensity, yielding more accurate scores; Andromeda \cite{cox2011andromeda}uses a binomial distribution as the basic scoring function.

Despite the database searching engine working well, these algorithms still have enormous room for improvement. Many researchers proposed the methodologies to re-score the PSM after using the database-searching process, a method called post-processing. Various publications focus on the post-processing of PSM scoring algorithms with two main methods. Many works use machine learning method to re-score the PSM: Percolator \cite{kall2007semi} trains semi-supervised supporting vector machines (SVM) by the features extracted from PSMs; Q-ranker \cite{spivak2009improvements}, based on Percolator, expands the feature set to train the SVM; Nokoi \cite{gonnelli2015decoy} splits the target and decoy data sets and uses logistic regression to train a post-process model. The other popular approach to improve PSM identification is to re-score the PSM with statistical models: Iprophet uses the expectation-maximization algorithm to build statistical models iteratively.

Recently, as deep learning has become efficient and effective in the pattern recognition area, some researchers propose deep learning architectures in the proteomics area. DeepNovo \cite{tran2017novo} implements a sequence-to-sequence architecture, which combines a CNN-based ion detection model with an RNN-based peptide sequence decoding model in order to generate De nova peptide sequences from tandem mass spectrometry data. DeepMatch \cite{schoenholz2018peptide} constructs a deep neural network that uses Bi-LSTM as a mass spectrum encoder and uses CNN as a fragment ion detection model to build a post-processing model for PSM identification for a single organism.

The deep learning model application is feasible to detect patterns between the mass spectra and fragment ions of peptides. Since the increase of the resolution for the mass spectra results in a sizable increase in the complexity of metaproteomics, we hope the deep learning model's power helps detect PSM's potential pattern.

\section{Method}
\begin{figure} % picture
    \centering
    \includegraphics{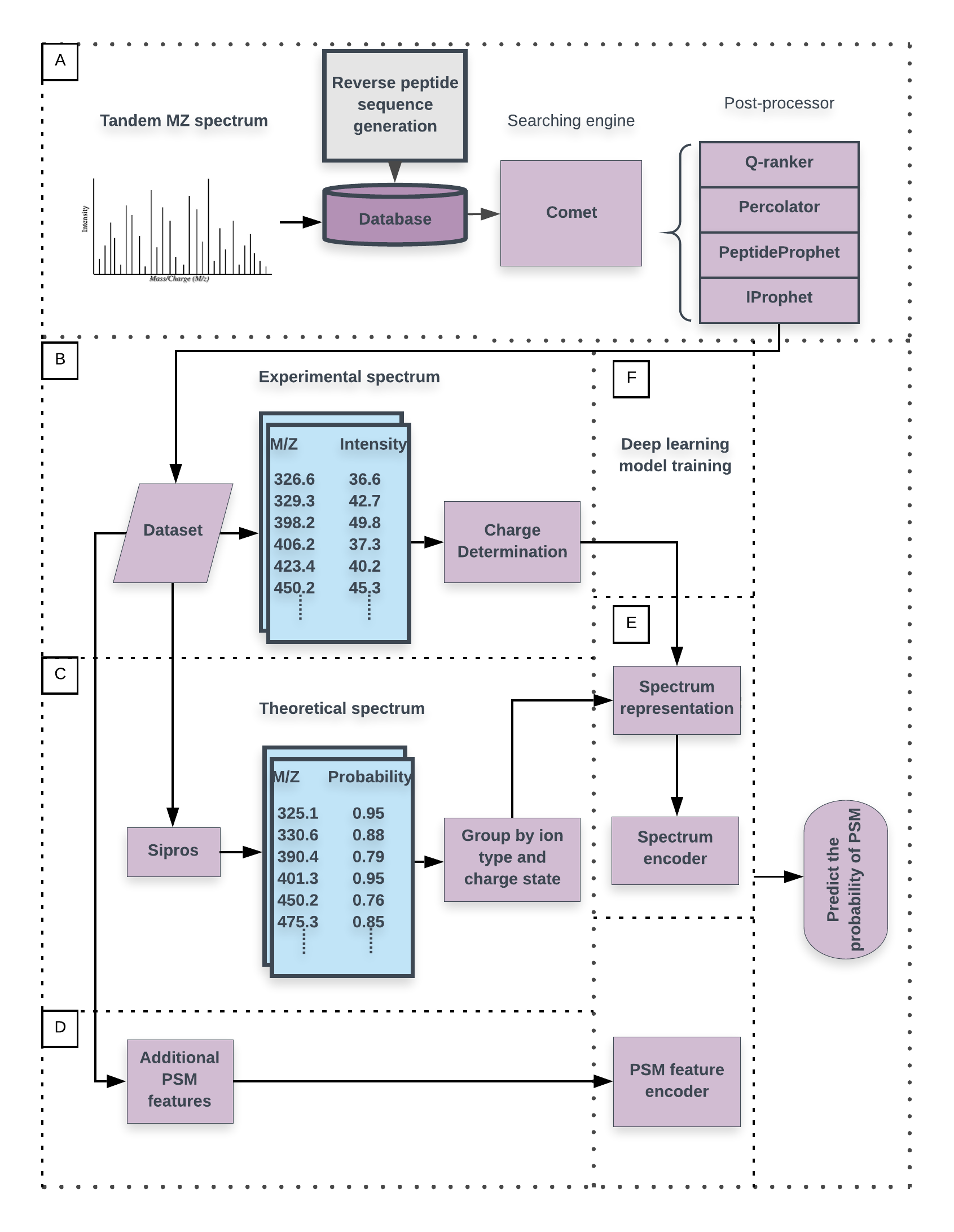}
    \caption{The workflow of building DeepFilter}
    \label{fig:workflow}
\end{figure}

We will discuss the details of our model in this section, which we refer to as DeepFilter. DeepFilter is consists of six major components,sub-figures A to F in the workflow \ref{fig:workflow} shows the relationship among the components mentioned following.: representative training PSM candidates selection and data set construction(A), charge detection for experimental spectrum (B), isotopic distribution generation (C), representation of spectrum and 11 PSM extra features (D) (E), and CNN based deep learning model (F). In brief, we firstly used a database-searching engine, Comet, and post-processing tools to investigate the tool which has better PSM identification performance; then, we constructed a representative training data set by controlling a threshold of posterior error probability. After that, the training data set was processed by charge detection algorithm and isotope distribution generation algorithm and transferred into spectrum representation to be fed into the CNN model, which is referred as spectrum encoder. The other features we used are attributes extracted from the corresponding PSM candidates; these 11 extra PSM features were fed into a fully connected layer, which referred to the PSM feature encoder. The above two encoders are the cores of our deep learning model. We will explain the details for each component in the following.

%Our purpose is to develop a post-processor based on deep learning for the existing database-searching engine to improve PSM identifications. Our model is consist of five major parts, which are shown as sub-figures in the workflow \ref{fig:workflow}:  - charge detection, representative training samples selection, isotopic distribution generation, spectrum encoder and PSM feature encoder, CNN-DNN based deep learning model.  The workflow of our work is shown in Figure 1. In brief, we firstly used a database-searching engine, Comet, and post-processing tools to get a set of top-scored PSMs, then we constructed a training dataset with high-quality by setting a threshold of posterior error probability. We fed the dataset into a CNN-based deep neural network architecture to train a classification model. The model was used as the post-process after PSM candidate's selection through database searching engine and  probability predicted by this model is regarded as the score of each PSM candidate for the later filtering.  In the rest of this section, we will discuss the details of data collection, input construction, and the architecture of DeepFilter.

\subsection{Data sets construction}
Traditional methods to identify PSM is to compare the similarity between the peptide sequence and experimental spectra using mathematical and statistical methods. Moreover, they use the target-decoy search strategy \cite{elias2007target}, which involves reverse protein sequences as decoys into the protein database, to select the confident PSMs; This strategy is to estimate false discovery rates (FDR) and regard the FDR as a threshold to filter the most confident PSMs in protein identification. For this process, most wide-used tools contain the function to initially digest the protein sequences and get the most similar peptide sequence for each specific scan. In our experiment, we firstly used Comet to collect a set of top-scoring PSM candidates for each scan, and re-score these PSMs by several wide-used post-process algorithms, which includes: Percolator \cite{kall2007semi}, Q-ranker \cite{spivak2009improvements} and PeptideProphet \cite{shteynberg2011iprophet} and IPropeht \cite{shteynberg2011iprophet}. After investigating these post-process tools, we select the algorithm that identifies most PSMs using the target-decoy strategy as the post-processor, which is the Percolator in our experiment, to generate the training data set.

There are seven data sets in our experiments, three data sets are metaproteome from a marine microbial community \cite{bryson2016proteomic}, and three data sets are from a soil community \cite{butterfield2016proteogenomic}; we also use an E.coli proteome to test if our model also fits the single organism. One data set of the marine microbial community is used to train the model, and other data sets are test data sets for the benchmark.

The training data was processed as the workflow described in the first paragraph. Go through the first pass by Comet and second pass by Percolator and collected the top-5 scored PSM candidates for each scan. After removing the PSM candidates whose posterior error probabilities are more than 0.93, we finally obtained a training data set containing 926,253 PSM candidates. For the annotation, the top-1 PSM candidates, which are target PSMs, will be labeled as positive PSMs, the top-1 PSM candidates, which are decoys, and the rest 4 PSM candidates of top-5 PSM candidates will be labeled as negative PSMs.

%We used Sipros to get the isotope distributions for the peptide sequences of these PSM candidates, and group each distribution by fragment charge and ion type (in our experiment, for the charge state, we just consider charge which equals to 1, 2 and 3; As for ion type, we just consider B-ion and Y-ion). After that, we extracted the intensity and probability for each PSM candidate and construct the data input matrix for later training process. Combined with experimental mass spectrum data process, here we get a original PSM candidate representation.

\subsection{Charge detection of experimental mass spectra}
The intensity distribution of MS2 files is mixed by different fragment ions of the peptide sequences. To capture patterns in different fragment ions, we firstly deconvolute the mass spectrum by applying a charge detection algorithm for experimental mass spectra and detect the charge for fragment ions. We leverage MaxQuant to process the ms2 files first to get the most redundant M/Z - intensity pairs; this kind of pairs are recorded in an APL format file. We reconstruct the experiment mass spectra by identifying the charge state for the most abundant peaks and representing them into a charge - m/z mapping dictionary for later process. The detail of the algorithm is described in the following.

\begin{algorithm}
\DontPrintSemicolon
  
  \KwData{each scan MS2 file and apl format file}
  \KwResult{Plain text file which contains the charge - m/z and intensity (most redundant mappings) for each scan}
  Initialization: $W_h$: weight of hydrogen; $W_n$: weight of neutron \;
                  
  \For{scan in MS2 file}
  {
    Initialization: $Group_i$, $i$ means the detected charge for the peak\;
    \For{m/z in scan}
    {
        
        \For{ charge in range (1 to 3)}
        {
            $detect_mz=mz * charge - (charge - 1) * W_h + j * W_n$\;
            \tcp*{Discharge and consider the isotope}\;
            \If{Search ($detect_mz$, apl file)  is  True) }
            {
                m/z-intensity $\in Group_{charge}$\;
            }
            \Else
            {
                m/z-intensity $\in Group_{None}$\;
            }

        }
   }
   WriteOut(scan, Groups, Plain text)\;
  }
\caption{Charge detection algorithms}
\end{algorithm}

\subsection{Isotope distribution generation}
We used Sipros \cite{hyatt2012exhaustive} to get the isotope distributions for the peptide sequences of the training PSM candidates. For each peak, we group each distribution by fragment charge and ion type (in our experiment, for the charge state, only the charge which equals to 1, 2, and 3 is considered; As for ion type, we only consider B-ion and Y-ion), and we controlled the cumulative isotopic abundance to be less than 98\%. After that, for each PSM candidate, we combine its isotope distribution and experimental mass spectrum after the charge detection process to get the spectrum representation for the later training process.

\subsection{Representation of spectrum and 11 PSM features}
\begin{figure} % picture
    \centering
    \includegraphics[width=180mm]{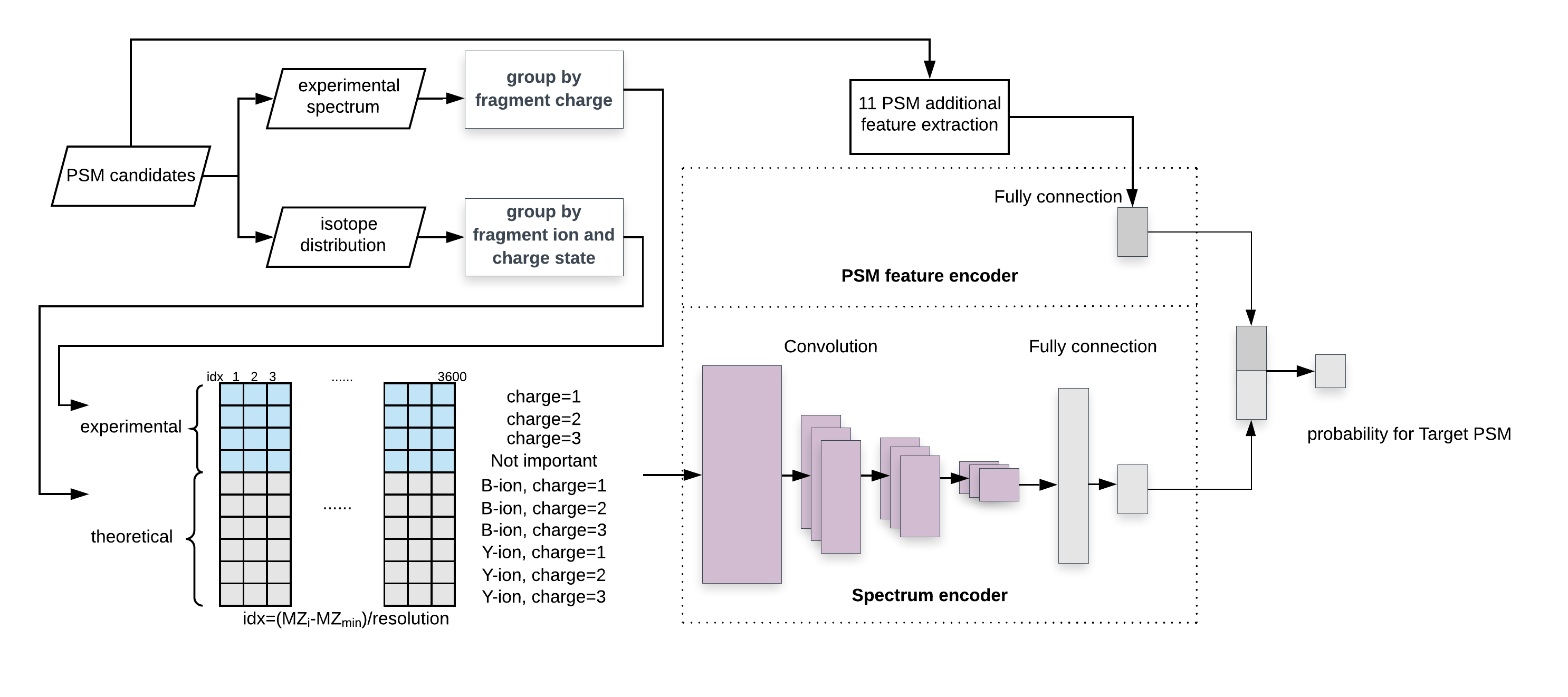}
    \caption{Data pre-processing and architecture}
    \label{fig:architecture}
\end{figure}

%Since our goal is to use deep learning  to detect the different pattern between high-scoring target and decoy PSM candidate compared with the same scan of experimental MS data, we tried to group the experimental and theoretical M/Z – intensity pairs by different charge state and ion type, and construct the input matrix by using the M/Z values to direct the matrix index and filling it with corresponding intensity or probability.\\
As the Figure \ref{fig:workflow} part B shows, a tandem mass spectrum always consisted of pairs of M/Z and intensity; for the theoretical spectrum (Figure \ref{fig:workflow} part C, the isotopic distribution for different fragment ions is presented as the theoretical M/Z and the abundance probability of the isotope. After the fragment charge detection for the experimental mass spectrum and the calculation of isotopic distribution for PSM candidates, which is shown as the Figure \ref{fig:architecture}, we get four groups of the experimental spectrum (group by three charge states and the peak whose charge state is not detected) and six groups of isotopic distribution (group by three charge states and two types of ions). In DeepFilter, to integrate the intensity of spectrum and isotopic distribution, the isotope's intensity and abundance probability are discretized. Then, combined with the identification of charge and ion type for the fragment ions, we constructed the spectrum representation matrix to fed into the CNN model. The representation of 11 extra PSM features is calculated after database-searching, and this representation is encoded with a fully connected layer. The details of these two representation is introduced in the following.\\
\paragraph{Spectrum representation}
Our spectrum representation is a matrix constructed by peaks with the charge from experimental mass spectra and fragment ions with charge state from isotope distributions. We encoded the spectrum representation by using the matrix index to indicate the M/Z, ion type, and charge state;  The details are shown in figure \ref{fig:architecture}. For each PSM candidate, we use 0.5 Da as a resolution parameter and regard the M/Z from 100 Da to 1900 Da and drop the rest. In this way, we initial an 8*3600 matrix with 0, then scan the sorting M/Z value lists group by group (four groups for experimental mass spectrum and six groups for isotopic distribution) for the specific PSM candidate. During the scan process, for each M$_i$/Z$_i$ pair, we calculate the index by the equation \textit{index =(M$_i$-M$_{min}$)/resolution}, then using the intensity to fill the matrix for the corresponding index in the first three rows within fragment charge equal to 1,2,3, and using the intensity of the  M$_i$/Z$_i$ pair whose fragment charge is not identified to fill the fourth row of the matrix; then using the abundance probability of the isotope to fill the rest rows of a matrix by the different combination of charge state and ion type (charge=1, Y-ion; charge=1, B-ion; charge=2, Y-ion; charge=2, B-ion; charge=3, Y-ion, charge=3, B-ion). After L2 normalization, the matrix will be fed to train the classification model. \\
\paragraph{Representation of 11 extra PSM features}
For each PSM candidate, we also extracted 11 features based on Comet for the later classification architecture. These 11 additional features are determined after investigating the weight of each feature used by Percolator and Q-ranker. The features are shown in Table \ref{tab:table}.  

\begin{table}
 \caption{11 extra PSM features used in Deep Filter}
  \centering
  \begin{tabular}{ccc}
    \toprule                 
    1 & Xcorr           & Cross correlation between calculated and observed spectra \\
    2 & $\Delta C_n$  & Fractional difference between current and second best XCorr     \\
    3 & $\Delta C_n^l$ & Fractional difference between current and third best XCorr      \\
    4 & Mass       & 	The observed mass [M+H]$^+$  \\
    5 & $Delta M$           & The difference in calculated and observed mass\\
    6 & $abs(\Delta M)$  & The absolute value of the difference in calculated and observed mass  \\
    7 & pepLen & The length of the matched peptide, in residues\\
    8 & enzInt       & The length of the matched peptide, in residues\\
    9-11 & charge 1-3       & Three Boolean features indicating the charge state\\
    \bottomrule
  \end{tabular}
  \label{tab:table}
\end{table}

\subsection{Model architecture}

We constructed a deep learning architecture that includes two encoders. The one consists of 4 convolutional layers and two fully connected layers; this encoder is called spectrum encoder. Spectrum encoder is used to detect the pattern of the top-scoring spectrum with different fragment ions and charge state.  The other encoder is consists of a single fully connectional layer as an extra feature pattern recognizer, which is called PSM feature encoder in later part. The input for the spectrum encoder is the spectrum representation, and the input for the PSM feature encoder is 11 extra PSM features. Then we concatenated the output from these two encoders and went through a fully connected layer to get a probability that if the PSM candidate is a target or decoy PSM. The architecture of our model is described in Figure \ref{fig:architecture}, and the detail of each encode will be introduced in the following.

\paragraph{Spectrum encoder}
The spectrum encoder is a deep neural network model with four convolutional layers and 2 two fully connected layers. To capture the high-scoring Target PSM candidate pattern, we used small kernel sizes to recognize the fragment ion. For the four convolutional layers, we use 16 kernels for each layer, and in each convolutional layer, we use different sizes of kernels, which are (4,7), (2,11), (2,9), (2,9) to capture the unsupervised features given by spectrum. We used max-pooling with (1,3) kernel size to capture the most weighted feature because of the high-dimension after each convolution operation. To speed up the process and avoid the over-fitting issue, we also apply batch-normalization for each convolutional layer and add a dropout layer after the last convolutional layer, whose disabled probability is 0.5. In the first fully connected layer, the input dimension is 3076, and the hidden units are 1024. For the second fully connected layer, the input dimension is 1024, and the output vector is 512, which is activated by ReLU function, and used as the representation of spectrum encoder for the later classification model.\\

\paragraph{PSM feature encoder}
For each PSM candidate, several PSM related properties could be calculated after the process of Comet. As described above, we fed the 11 features obtained by Comet for each sample into a single fully connected layer, the input dimension for this layer is 11, and after activation by ReLU function, a 512-dimension vector is an output. This vector was used as the representation of PSM feature encoder.\\

\paragraph{Scoring model}
The representations from the spectrum encoder and PSM feature encoder were then concatenated together into a 1024-dimension matrix and fed into another fully connected layer with the softmax activation function. The output will be the probability from 0 to 1 to predict if a PSM candidate is a target PSM.\\ 

\paragraph{Loss function}
As described in the section Data set construction, the data set is annotated as positive or negative, which means the scoring model is a binary classifier. However, the label is given by scoring the PSM candidate with Comet, but not a real label. To enhance the model to detect more target PSMs, we apply an advanced cross-entropy loss function by involving the posterior error probability (PEP) parameter after filtered by Percolator. We regarded the \textit{p$_i$} as the correct probability calculated by PEP and regarded \textit{t$_i$} as the prediction probability. Furthermore, we can call the loss function Equation \ref{eqn:loss} to update the weights for the model.\\

\begin{equation}
\label{eqn:loss}
Loss=-\sum[p_{i}\log t_{i}+(1-p_{i})\log (1-t_{i})]
\end{equation}

\section{Experiment and Results}
\subsection{Experiment design}
We evaluated the performance of DeepFilter compared with the other four post-process tools, which is state-of-art in different Shotgun proteomics tasks. The five algorithm includes: Percolator \cite{kall2007semi}, Q-ranker \cite{spivak2009improvements}, PeptideProphet \cite{nesvizhskii2003statistical} and IProphet \cite{shteynberg2011iprophet}. Percolator, Q-ranker, and PeptideProphet are all using Comet as the pre-process searching algorithm. For IPropeht, this algorithm uses the PeptideProphet to processed the PSM candidates from the database-searching engine. Since the IProphet algorithm uses peptide and protein level features to improve the result of the PSM level, which causes the filtering of that two levels is not independent of PSM level filtering. In this way, we also experiment with disabling 5 statistic models when applying Iprophet to evaluate the performance from different perspectives.\\

The data sets to be evaluated are the six testing data sets from marine and soil microbial communities, as we introduced in the method section. The consistency of the data sets is shown in Table \ref{tab:comsistency}. First, We use Comet as the searching engine to get a set of the confident peptide sequence. Then we post-processed the peptide sequences by the benchmark post-processors above. To measure each post-processor's performance, we calculated the amount of PSM, peptide, and protein. For every spectrum, only the PSM candidate with the highest score was regarded as the PSM for this spectrum; then, we set different FDRs as the threshold to compare the post-processor's performance. The FDR calculation equation we followed is shown in Equation \ref{eqn:fdr}. In this equation, \#\textit{Target} is the amount of target PSMs and the \#\textit{Decoy} is the amount of decoy PSMs.\\

\begin{equation}
\label{eqn:fdr}
FDR = \frac{\#Target}{\#Decoy}
\end{equation}

%We also filtered existing post-process algorithm which show good performance in Shotgun proteomics within these five complex microbial communities and the single organism. The five algorithm includes: Percolator \cite{kall2007semi}, Q-ranker \cite{spivak2009improvements}, IProphet \cite{shteynberg2011iprophet}. Percolator and Q-ranker are all using Comet as pre-process searching algorithm. For IPropeht, this algorithm should be initialized by PeptideProphet algorithm, and it contains 5 models to refine the initial probabilities for peptide identification, so we also test the PeptideProphet and disable the application for 5 models.

\begin{table}[]
    \label{tab:comsistency}
    \centering
    \begin{tabular}{lccccccc}
    \hline
         &\textbf{marine 1}&\textbf{marine 2}&\textbf{marine 3} &\textbf{soil }&\textbf{soil 2}&\textbf{soil 3}&\textbf{ecoli} \\ \hline
         \\
         \# of spectra& 138682 & 143344&127075&804404&1010954&843212& 1439140\\ \hline
    \end{tabular}
    \vspace{2mm}
    \caption{The consistency of data sets}
\end{table}

\subsection{Performance Comparison among Deep filter and other Post-process tools}
DeepFilter is applied to five data sets of complex microbial communities. Our method compared peptide identification results with Comet database-searching algorithm and the other four existing post-process algorithms - Percolator, Q-ranker, PeptideProphet, and IPropeht - at PSM, peptide, protein level within the threshold equals to 1\%. The result is shown in the Table \ref{tab:metaproteom}, the row name IProphet-NON represents the experiment to disable all the statistic models; the bold entry is the best result for the specific data set, and the underlined entry is the second best. We also applied the model in the single organism - E.coli at three different FDR levels equal to 1\%. The result is shown in Table \ref{tab:singleOrganism}.\\
\begin{table}
\label{tab:metaproteom}
\centering
\begin{tabular}{lccccc}
\hline  & \textbf{Marine 2} & \textbf{Marine 3}& \textbf{Soil 1}& \textbf{Soil 2}& \textbf{Soil 3} \\ \hline
\\
 & \multicolumn{5}{c}{PSM identification at PSM level within FDR 1\%}\\
 \\
 Comet &31822&38490&79505&75693&72454\\
Percolator & 34741& 41714& 88037& 84623&81331\\
Q-ranker & 33899& 40832& 86433& 82773&79006\\
PeptideProphet& 30670& 37072& 73821& 71281&68067\\
IProphet & \uline{38476}& \uline{44588}& \textbf{95501}& \textbf{94090}&\textbf{89959}\\
IProphet-NON & 30846& 37304& 75360& 73331&70121\\
Deep filter & \textbf{38927}& \textbf{44664}& \uline{92221}& \uline{89465}&\uline{86809}\\ \\\hline
\\
 & \multicolumn{5}{c}{Peptide identification at peptide level within FDR 1\%}\\
\\
 Comet &22004&25085&26068&23500&20423\\
Percolator & 24150& 27522& 29304&26989&23275\\
Q-ranker & 23589& 26674& 29163&26116&23673\\
PeptideProphet& 21597& 24653& 25288&23478&19863\\
IProphet & \uline{25787}& \uline{28539}&\textbf{31260}&\uline{28741}&\textbf{25160}\\
IProphet-NON & 21696& 24661& 25403& 22775&19922\\
Deep filter & \textbf{26582}& \textbf{29300} & \uline{30111}& \textbf{28968} &\uline{25006}\\ \\\hline
\\
 & \multicolumn{5}{c}{Protein identification at Protein level within FDR 1\%}\\
\\
 Comet &7033&7457&6938&6913&5644\\
    Percolator & \uline{7715}& \uline{8209} &\uline{7756}&\uline{7519}&6183\\
Q-ranker & 7617& 8151& 7684&7498&\uline{6387}\\
PeptideProphet& 7039& 7354& 6821&6848&5473\\
IProphet & 7108 & 7400& 6850&6928&5544\\
IProphet-NON & 5375& 7433& 6819& 6879&5577\\
Deep filter & \textbf{8313}& \textbf{8851} & \textbf{8069} & \textbf{8041} &\textbf{6976}\\ \\\hline

\end{tabular}
\vspace{2mm} 
\caption{\label{font-table}Identification performance using five real-world metaproteomes at FDR 1\%}
\end{table}

\begin{table}
    \label{tab:singleOrganism}
    \centering
    \begin{tabular}{lccc}
    \hline & PSM & Peptide & Protein \\ \hline
    \\
       Comet  & 504466 &30944&2147 \\
       Percolator &\uline{508017}&31179&\uline{2167} \\
       Q-ranker & 507425 &30829&2165\\
       PeptideProphet &493686&30014&2060\\
       IPropeht &498810&\uline{31201}&2041\\
       IPropeht-NON &495353&30037&2057\\
       Deep filter &\textbf{509961}&\textbf{31247}&\textbf{2172}\\ \\ \hline
    \end{tabular}
    \vspace{2mm}
    \caption{\label{font-table}Identification performance using E.Coli data set at FDR 1\%}
\end{table}
The results in Table \ref{tab:metaproteom} shows the peptide identification amounts at a different level when accepting the false discovery rate equals to 1\%. From the table, the comparison between different post-processor based on Comet could be observed. At the PSM level and peptide level, our model and IProphet always achieve the top-2 PSM detection amount: our model outperforms at the PSM level in marine data sets, and IProphet outperforms at the peptide level in soil1 and soil3 data sets. However, although the IPropeht shows better performance at PSM and peptide level, the PSM and peptide level increase does not help improve the protein identification. That may be caused because the Independence between PSM level and protein level means Iprophet uses some features at the protein level to improve the PSM identification. Our model always outperforms at the protein level at FDR 1\%; the improvement fraction is shown in Table 5. Second, the best column shows the second-best target protein detection amount at the protein level within FDR 1\%. Compared with the second-best algorithm, our model mostly achieves more than 5\% improvement except in the soil1 data set, but for the soil1 data set, there is also a slight improvement. For the single organism E.coli, DeepFilter also ac hives comparative protein identification performance among the bench-marking post-processor in different peptide identification levels within FDR equals to 1\% 

\begin{table}
\label{tab:comparison}
    \centering
    \begin{tabular}{lccc}
    \hline & Second best & Deep filter & improvement \\ \hline
    \\
       marine2  & 7715 &8313&7.19\% \\
       marine3 &8209&8851&7.25\% \\
       soil1 & 7756 &8069&3.88\%\\
       soil2 &7519&8041&6.49\%\\
       soil3 &6387&6976&8.44\%\\\\ \hline
    \end{tabular}
    \vspace{2mm}
    \caption{\label{font-table}improvement fraction for complex microbial communities at Protein level within FDR 1\%}
\end{table}

\subsection{Performance comparison at different FDR}
\begin{figure}
\label{fig:PSM}
    \centering
    \subfigure[marine2]{\includegraphics[width=80mm]{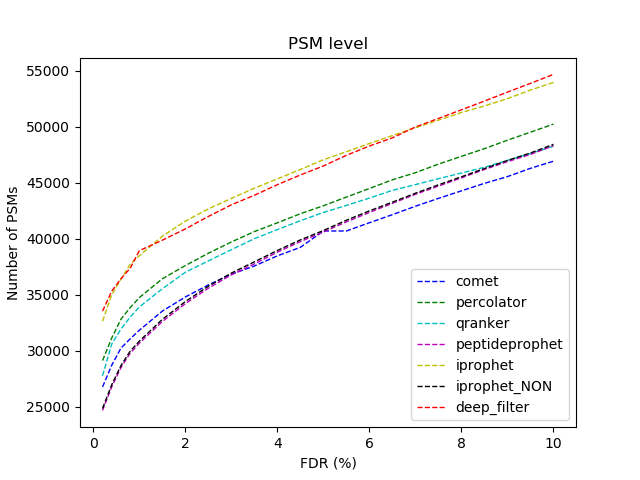}}
    \subfigure[marine3]{\includegraphics[width=80mm]{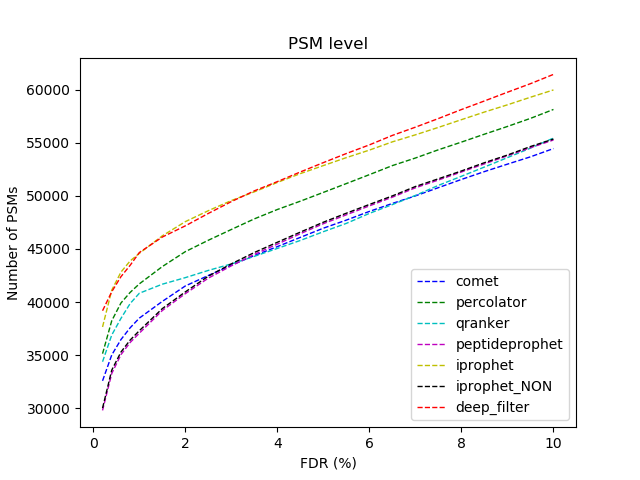}}
    \subfigure[soil1]{\includegraphics[width=80mm]{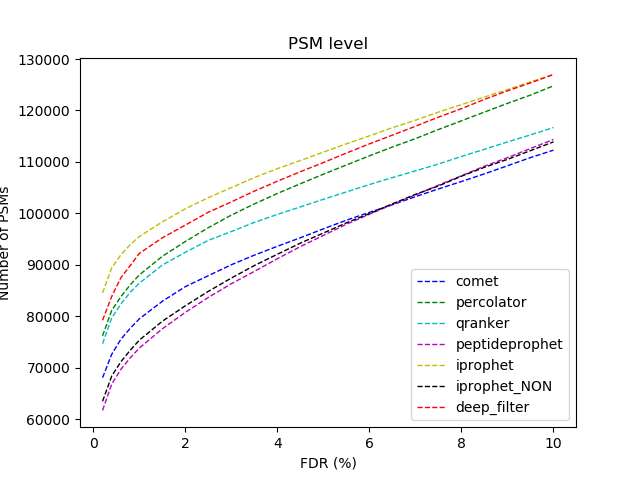}}
    \subfigure[soil2]{\includegraphics[width=80mm]{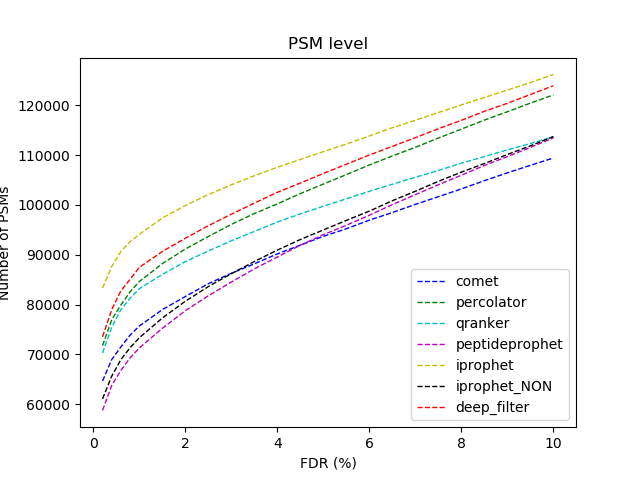}}    \subfigure[soil3]{\includegraphics[width=80mm]{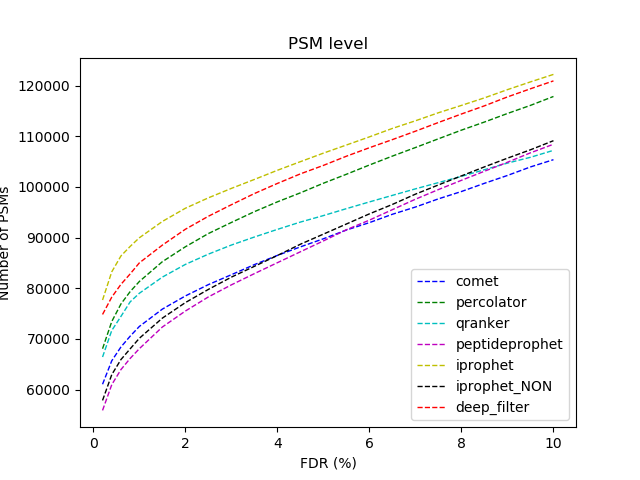}}
    \subfigure[ecoli]{\includegraphics[width=80mm]{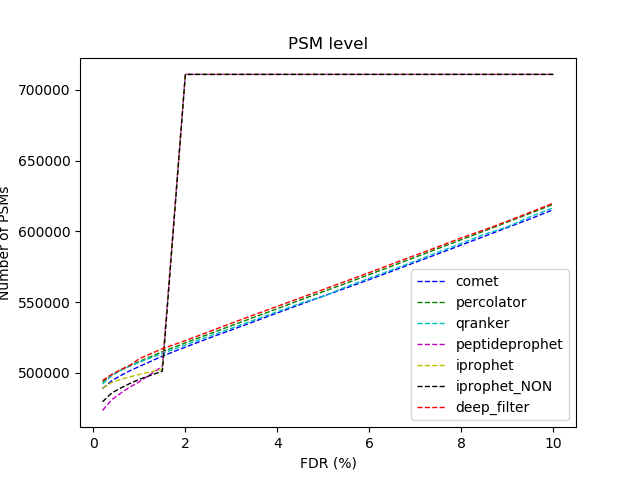}}
    \caption{PSM performance within different FDR.}
\end{figure}

\begin{figure}
\label{fig:peptide}
    \centering
    \subfigure[marine2]{\includegraphics[width=80mm]{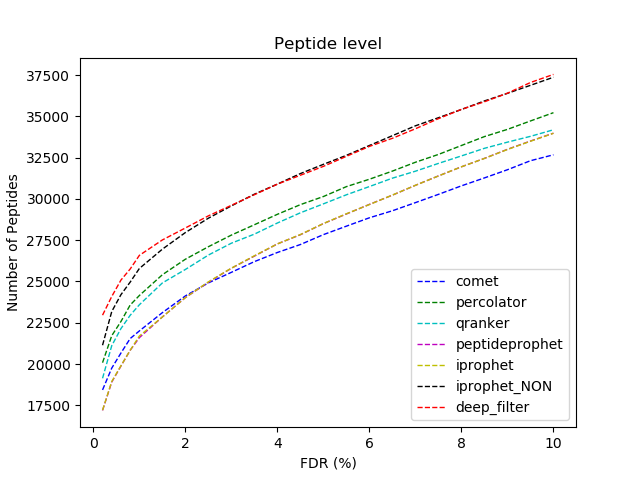}}
    \subfigure[marine3]{\includegraphics[width=80mm]{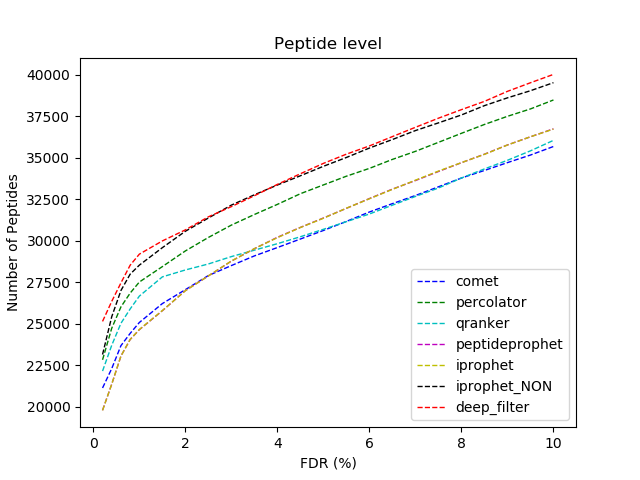}}
    \subfigure[soil1]{\includegraphics[width=80mm]{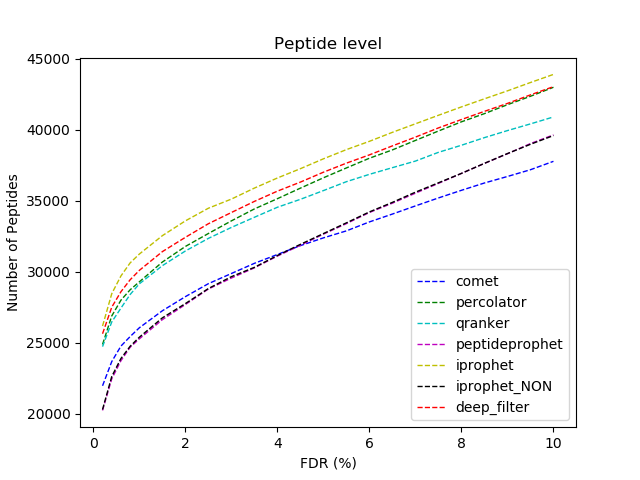}}
    \subfigure[soil2]{\includegraphics[width=80mm]{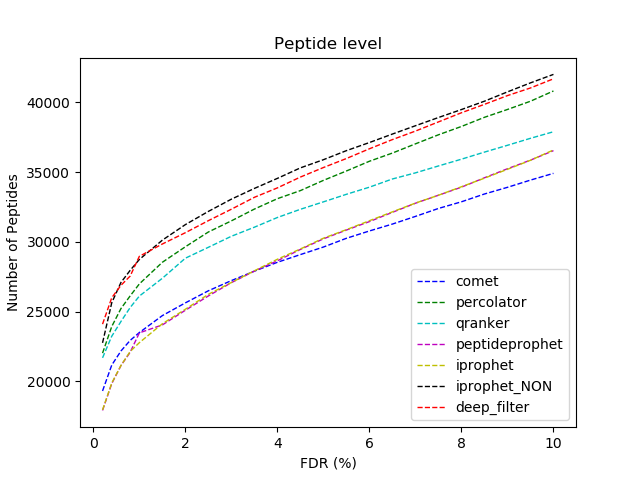}}    \subfigure[soil3]{\includegraphics[width=80mm]{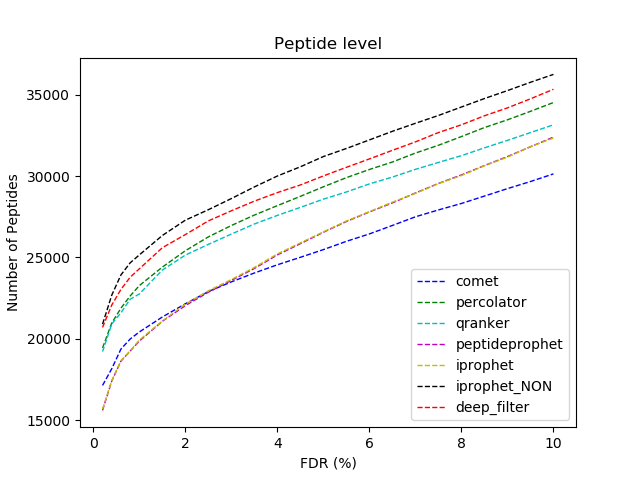}}
    \subfigure[ecoli]{\includegraphics[width=80mm]{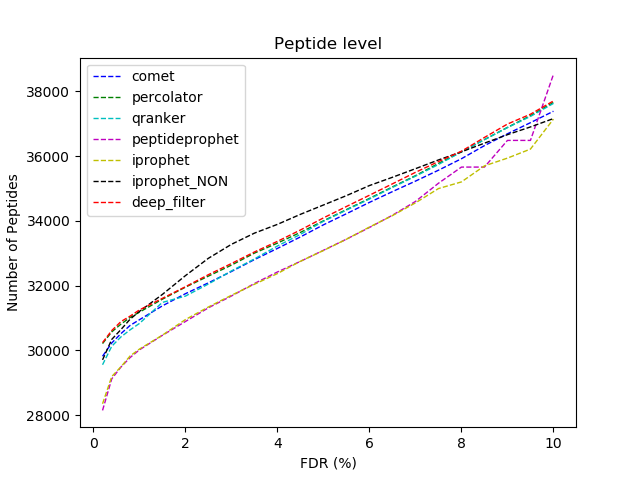}}
    \caption{Peptide performance within different FDR.}
\end{figure}

\begin{figure}
\label{fig:protein}
    \centering
    \subfigure[marine2]{\includegraphics[width=80mm]{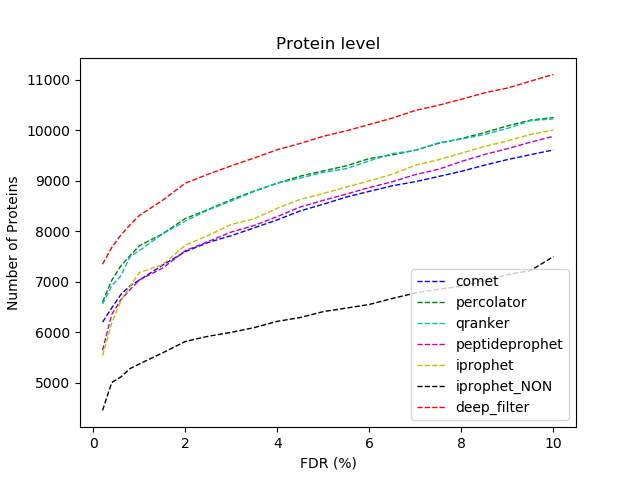}}
    \subfigure[marine3]{\includegraphics[width=80mm]{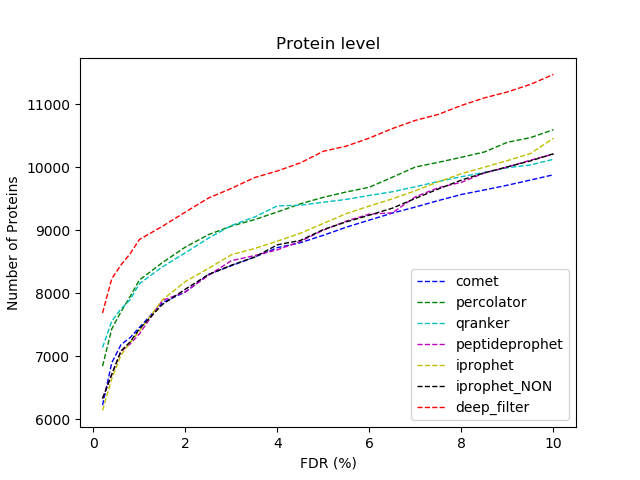}}
    \subfigure[soil1]{\includegraphics[width=80mm]{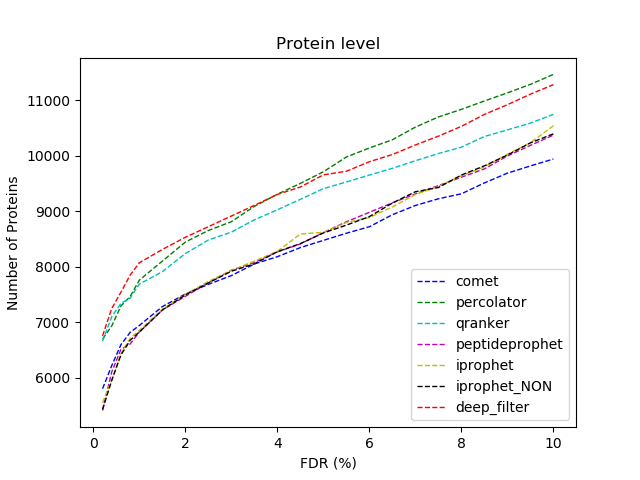}}
    \subfigure[soil2]{\includegraphics[width=80mm]{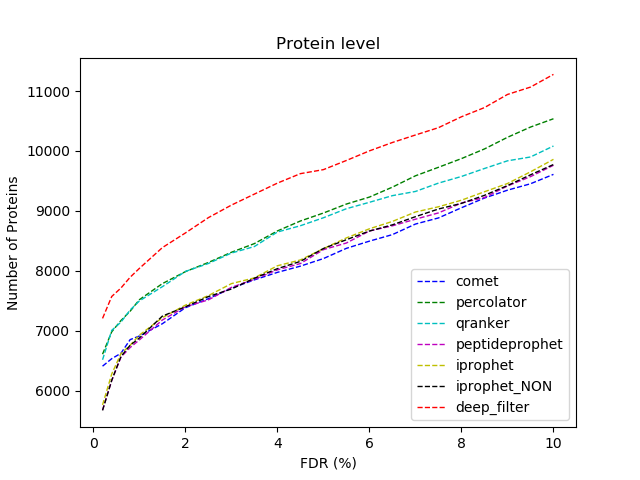}}    \subfigure[soil3]{\includegraphics[width=80mm]{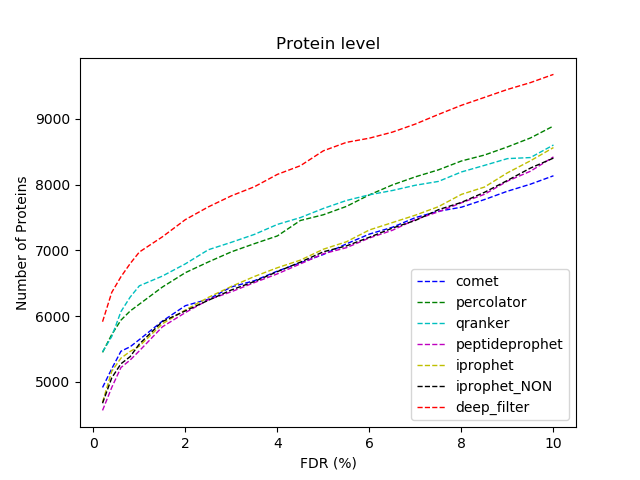}}
    \subfigure[ecoli]{\includegraphics[width=80mm]{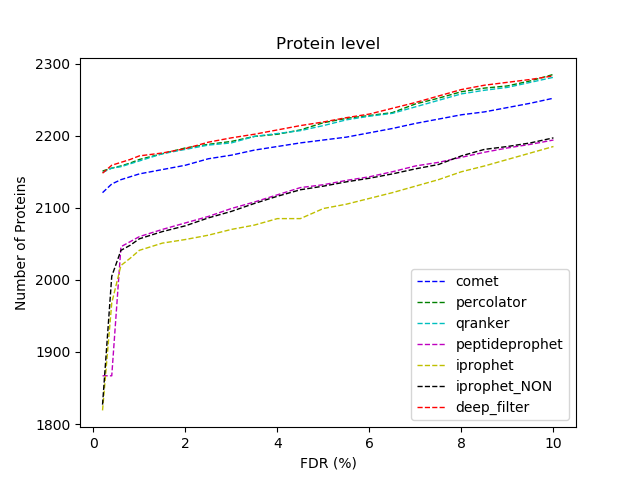}}
    \caption{Protein performance within different FDR.}
\end{figure}

We adjust the FDR from 0.2\% to 10\% to test the fraction of our model compared with other database-searching engine and post-process algorithms. 
Figure \ref{fig:PSM}, \ref{fig:peptide}, \ref{fig:protein} shows the experiments results, which conducted using the data sets describe above to identify peptide within different FDR value (from 0.2\% to 10\%) at PSM, peptide and protein level. For complex microbial communities, generally, IPropeht and Deep filter always stand at the top-2 position at PSM and Peptide level; IPropeht especially has a significant increment than any other data sets in the soil community. However, relying on the protein level feature to build a statistic model, Iprophet does not achieve a good performance at the protein level. For protein level, Deep filter, Percolator, and Q-ranker stand at top-3 positions within all acceptable FDR. Deep filter always outperforms except in soil1 data set from FDR 5\% to 10\%, considered of the training data set, and consistent with the marine microbial community, this situation may be caused by the lack of representative training samples. In this way, some potential patterns are not learned in the training process. For the desperately set in different communities, fine-tuning of our model within new data set may help to gain a post-processor with good performance just within a short training time. For the single organism, from Figure \ref{fig:PSM}, \ref{fig:peptide}, \ref{fig:protein}, the result of the E.coli data set shows our model can achieve a significant similar result with the best post-processor in the baseline. That is, DeepFilter can also be applicable in single organisms.

\section{Discussion}
There is much work to study how to improve peptide identification after database-searching with different machine learning methods. Percolator \cite{kall2007semi} and Q-ranker \cite{spivak2009improvements} using SVM algorithm; The PeptideProphet\cite{keller2002empirical} apply expectation-maximization (EM) to estimate the peptide identification probability. However, most of them use semi-supervised fashion to construct and train the data sets; They select positive PSMs by the control of FDR, and use the set of positive PSM combined with all the decoy PSMs to train the machine learning models, and applied that models to the whole data set and re-score the PSM candidates. In this way, partial PSMs are seen data, and every time the users who post-processes and re-score a set of PSM candidates, they must retrain a new model. Within DeepFilter, the experiments show that the model is considerably generalized for the unseen dataset; the model can be used in another data set, which means the unsupervised feature learning of CNN recognizes potential patterns. \\ 

With the development of deep learning, more research works have been published to present how to apply a deep neural network to detect patterns from tandem MS data in the proteomics area. For example, DeepMatch \cite{schoenholz2018peptide} uses VGG-16 to recognize the patterns between tandem MS data and encoded amino acid sequence to improve the number of PSMs, and uses three well known single organisms (Yeast, Human, Mouse) to evaluate the model; Deep Novo \cite{tran2017novo} uses the sequence to sequence model constructed by CNN and LSTM to generate de novo sequences. \\

Compared to the above-existing algorithms, we trained a CNN-based deep learning model to extract potential information from MS2 data and peptide sequences from the database-searching engine and estimate PSM's probability in new data sets using that model. We did not apply a semi-supervised version to involve a subset to train the model each time we post-process the searching engine results; After a model is trained, we directly use the model without any fine-tune settings to estimate the peptide probability for new data sets. Compared with the DeepMatch PSM identification model, our model has a simpler architecture and less neural network parameters, making the model more feasible to deploy. Besides, we evaluate our model not just at the PSM level with a specific acceptable FDR value, the peptide level, and protein level inference is also tested. Our model can also achieve better performance in complex microbial communities; we focus more on metaproteomics while DeepMatch analyzes single organisms. Some ensemble approaches show brilliant PSM identification results, such as MSBlender \cite{kwon2011msblender} uses a probabilistic approach while Sipros Ensemble \cite{guo2018sipros} uses the logistic regression algorithm to integrate peptide identification from multiple data searching engines. Nevertheless, this is another research method, which is not comparable to our approach.

To mine the patterns and visualize the features learned by unsupervised feature engineering of our deep learning architecture, we adopted a class activation mapping (CAM) generation technique \cite{zhou2016learning} to help us interpret the learning decision of the CNN model. The CAMs show the most influential image region, which contributes to predicting a particular category. In our experiment, we apply this algorithm in our spectrum representation to visualize the patterns to predict a target PSM.

\begin{figure}
\label{fig:cam1}
    \centering
    \subfigure[]{\includegraphics[]{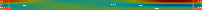}}
    \subfigure[]{\includegraphics[]{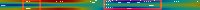}}
    \subfigure[]{\includegraphics[]{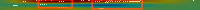}}
    \subfigure[]{\includegraphics[]{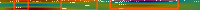}}    
    \caption{class activation mappings of target PSMs.}
\end{figure}

\begin{figure}
\label{fig:cam2}
    \centering
    \subfigure[]{\includegraphics[]{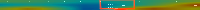}}
    \subfigure[]{\includegraphics[]{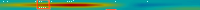}}
    \subfigure[]{\includegraphics[]{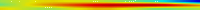}}
    \subfigure[]{\includegraphics[]{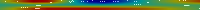}}    
    \caption{class activation mappings of decoy PSMs.}
\end{figure}

In Figure \ref{fig:cam1} and Figure \ref{fig:cam2}, we present four CAMs for target and decoy PSMs separately. In the figures, white points indicate mass spectrum data and corresponding intensity data in this position. Furthermore, the color represents the importance of the learning weight of CNN for this area. This region contributes more to predicting target PSM as the color of the background goes brighter; The rectangle with red lines is a region that shows an obvious ion matching pattern for MS data and theoretical data.

Figure \ref{fig:cam1} presents the CAM for target PSM, the sub-figures for different PSMs. We can see that the most significant part (red region) learned by CNN cover the fragment ion. Figure \ref{fig:cam2} shows the CAM for decoy PSMs; Sub-figures present different situation that the features CNN learned do not contribute to predicting a target PSM; Sub-figure (a) shows a low weighted region (blue region) to detect fragment ion matching in ions' region, sub-figure (b) have light class mapping, but it fails to cover the ion position, sub-figure (c) has a large part red region, but there is no efficient MS data, and sub-figure (d) has a dense mass spectrum data and CAM was covering, but they are not matching.


\bibliographystyle{unsrt}  
\bibliography{references}

\end{document}